# Hierarchically organized nanostructured $TiO_2$ for photocatalysis applications

F. Di Fonzo[1], C.S. Casari[1], V. Russo[1], M.F. Brunella[2], A. Li Bassi[1], and C.E. Bottani[1]

[1] Politecnico di Milano, Dipartimento di Chimica, Materiali e Ingegneria Chimica 'G. Natta', and NEMAS - Center for NanoEngineered MAterials and Surfaces, via Ponzio 34/3, 20133 Milano (Italy)
[2] Politecnico di Milano, Dipartimento di Chimica, Materiali e Ingegneria Chimica 'G. Natta', via Mancinelli 7, 20131 Milano (Italy)

E-mail: andrea.libassi@polimi.it

PACS: 81.07.-b; 81.16.Hc; 81.15.Fg

**Abstract**
A template-free process for the synthesis of nanocrystalline $TiO_2$ hierarchical microstructures by reactive Pulsed Laser Deposition (PLD) is here presented. By a proper choice of deposition parameters a fine control over the morphology of $TiO_2$ microstructures is demonstrated, going from classical compact/columnar films to a dense forest of distinct hierarchical assemblies of ultrafine nanoparticles (<10 nm), up to a more disordered, aerogel-type structure. Correspondingly, film density varies with respect to bulk $TiO_2$ anatase, with a degree of porosity going from 48% to over 90%. These structures are stable with respect to heat treatment at 400°C, which results in crystalline ordering but not in morphological changes down to the nanoscale. Both as deposited and annealed films exhibit very promising photocatalytic properties, even superior to standard Degussa P25 powder, as demonstrated by the degradation of stearic acid as a model molecule. The observed kinetics are correlated to the peculiar morphology of the PLD grown material. We show that the 3D multi-scale hierarchical morphology enhances reaction kinetics and creates an ideal environment for mass transport and photon absorption, maximizing the surface area-to-volume ratio while at the same time providing readily accessible porosity through the large inter-tree spaces that act as distributing channels. The reported strategy provides a versatile technique to fabricate high aspect ratio 3D titania microstructures through a hierarchical assembly of ultrafine nanoparticles. Beyond photocatalytic and catalytic applications, this kind of material could be of interest for those applications where high surface-to-volume and efficient mass transport are required at the same time.

## 1. Introduction

Nanostructured titanium oxide is a relevant material for a number of emerging energy and environmental technologies, like Dye Sensitized Solar Cells (DSSC), organic photovoltaics (OPV), photocatalysis for direct hydrogen production and environmental remediation [1-4]. Two main research pathways have been followed, so far, in order to increase the efficiency of the above-mentioned technologies. Several studies have been devoted to shifting the absorption limit of the material in the visible range, searching for an increase in reaction rate of photocatalysis and hydrogen production [5-7]. Another possible route is the maximization of the effective surface area as suggested by D.R. Rolison [8], underlying "the Importance of Nothing and the Unimportance of Periodicity" in catalytic nano-architectures, where "nothing" is the porosity at the nanoscale. The rationale beside these efforts resides in the basic argument that the efficiency of a surface mediated reaction increases with surface area. Ideally, one would like to have a full solar spectrum absorbing material with the highest specific surface area onto the smallest footprint. Photocatalysis is similar to photosynthesis in the conversion of photon energy to chemical, hence it seems reasonable taking inspiration from Nature to enhance it, at least regarding the issue of maximizing the surface area. As an example, a forest possesses a hierarchical and self-organized structure starting from the single tree with trunk, branches and leaves up to the entire tight forest assembly. This superstructure maximizes the gaseous exchange with the atmosphere maximizing leaves area, while at the same time optimizes leaves exposure to the sun light. In a similar way, a hierarchical structure with a multiscale organization and with a large and accessible surface area would be desirable for photocatalytic applications. Neglecting dispersion-on-support methods [9], the most suited approach is the assembly of nanoparticles, as small as possible, in a porous mesostructure. As far as titanium oxide is concerned, sol-gel synthesis of nanoparticles followed by sintering has been employed, generally yielding disordered films either with moderate porosity [10, 11], or in the form of an aerogel [12,13]. Ordered structures, with an engineered pore architecture, have been obtained by using templating agents [14-17]. Besides the complexity of this approaches, the necessary removal of the templating agent may cause severe dimensional changes and surface area reduction [16]. Very recently, there have been reports on the template-free fabrication and photocatalytic activity of hierarchically organized nanostructured $TiO_2$ [18-21]. Although the employed techniques are very powerful in producing powders composed by highly engineered particles, characterized by a complex nanostructure, no reports have been issued on the growth of hierarchical $TiO_2$ thin films on surfaces for the functionalization of large scale areas. Typical thin film deposition techniques, such as CVD, sputtering and Pulsed Laser Deposition (PLD), have also been tried with some excellent results in the doping issue but limited success in regards to surface area, since they tend to form essentially 2D films with limited surface area [22-25]. Few authors obtained columnar or sculptured thin films by sputtering [26], evaporation [27], or flame aerosol synthesis [28], with porosity determined by column diameter and spacing. In particular, Suzuki and Yang obtained high surface area $TiO_2$ and ZnO films showing enhanced activity by typical thin film techniques like Glancing Angle Deposition (GLAD) and MOCVD, respectively [27, 29]. Suzuki demonstrated an enhanced surface reaction efficiency of obliquely deposited $TiO_2$ thin films with variously shaped columns (e.g. zigzag, cylinder, and helix). He clearly showed the columnar thickness and spacing playing an important role in the enhancement of the effective surface area. Noteworthy examples of 3D film structure were demonstrated by Goossens et al. who obtained a fractal 'forest like' titanium oxide deposit by CVD [30]. In general, the active surface area of these films is limited to the sum of the exposed areas of individual columns or needles. Furthermore, inter-columnar spacing is hardly controllable to a large extent. We demonstrated, in the case of tungsten oxide [31], that PLD is a good candidate to address this issue thanks to its versatility in producing materials with defined structure and morphology. In a previous study [32], we performed comparative tests of photocatalytic activity of various surfaces, showing that PLD-produced $TiO_2$ films can be superior to anodized-annealed titanium surface and pure anatase powder.

In this communication we demonstrate the one-step, large scale and template–free production, by reactive PLD, of hierarchically organized nanostructured $TiO_2$. Our PLD bottom-up synthesis approach allows the assembling of ultra-fine nanoparticles (<10 nm) in a hierarchical superstructure that resembles a 'forest of trees', without any use of templates or pre-patterning over an area of more than 20 $cm^2$. By optimizing the PLD parameters, we were able to grow the single 'tree' structure and the entire 'forest' assembly, with tailored crystalline structure and morphology at the sub-micrometer scale. Noticeably, such architecture shows enhanced photocatalytic activity with respect to standard Degussa-P25 powder.

## 2. Experimental details

Pulsed Laser Deposition (PLD) of titanium oxide thin films was accomplished by ablating a pure Ti target with laser pulses from a KrF excimer laser (wavelength 248 nm, duration 10–15 ns, energy density 4 $J/cm^2$) in different dry air background pressures. PLD thin films were grown on both titanium and silicon substrates at room temperature. The effect of thermal treatment was investigated by 1 hour annealing at a temperature of 400 °C in air. X-ray thin film diffraction (Philips PW3020, Cu Kα radiation) and Raman spectroscopy (Renishaw InVia, excitation wavelength 514.5 nm) were used to characterize the oxide allotropic phases. Scanning Electron Microscopy images of the samples were acquired by a Field Emission SEM (Zeiss SUPRA 40). Photocatalysis tests were carried out by irradiating samples with a UV lamp (Xe 300W HAMAMATSU Super Quiet). Light was focused by a convergent lens; a quartz cell filled with water was used to absorb the IR component of radiation in order to avoid stearic acid sublimation. The radiation intensity was 0.2mW/cm$^2$. Standardized tests were performed by covering the sample surface with a controlled amount of stearic acid diluted in hexane. The mineralization degree was determined at several irradiation times by FTIR (Nicolet 510P) measurements. Concentration abatement was measured by monitoring the IR peak corresponding to the asymmetric stretching mode ($CH_2$ group) at 2923 cm$^{-1}$. Optical microscope observations were performed to evaluate the distribution of stearic acid crystals and their time-dependent degradation. In order to elucidate how different hierarchical organizations affect film activity, care was taken to deposit a known mass of $TiO_2$. Precise control of sample mass was pursued by careful calibration of the deposition rate employing a Quartz Crystal Microbalance (QCM) and HRSEM micrographs of the cross section of "ad hoc" samples. In this way, film density was calculated with respect to a bulk titanium oxide film of the same mass. Band gap energies were obtained by the *Tauc Plot* method from diffuse reflectance spectra acquired with a JASCO V570 (spectrum of Degussa-P25 powder was used as reference). The BET surface area was analyzed by using the nitrogen adsorption method in a Micromeritics Tristar 3000 (USA). An *ad hoc* sample, about 10 mg in weight, was prepared for reliable measurements. All the samples were degassed at 120 °C for 4 h prior to the nitrogen-adsorption measurements. The BET surface area was determined by using the adsorption data in a relative pressure ($P/P_0$) range of 0.05–0.3.

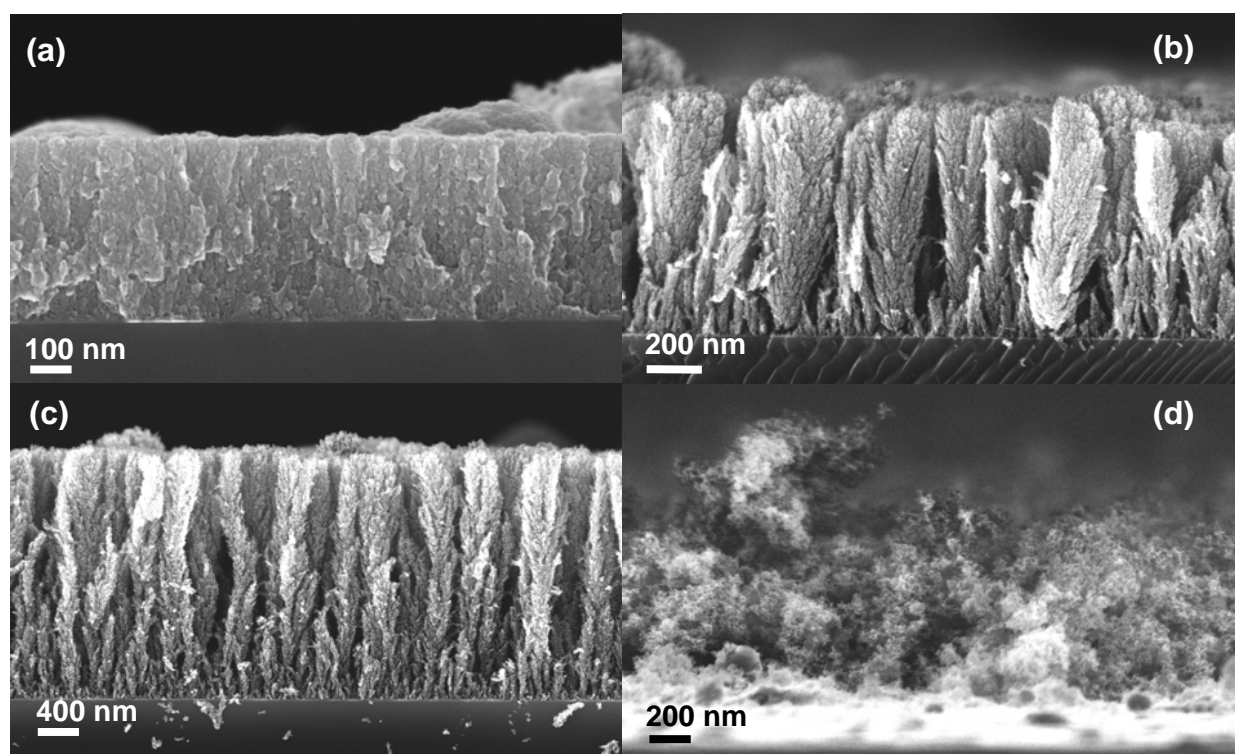

**Figure 1.** Cross sectional images of samples deposited at 10 Pa (a), 20 Pa (b), 40 Pa (c) and 100 Pa (d) after one hour annealing in air at 400°C.

## 3. Results and Discussion

Laser ablation in the presence of a reactive background gas pressure results in cluster formation in the ablation plume and thus in the production of nanostructured layers with a degree of porosity at the nano- and meso-scale, which increases the specific surface area [33-35]. By controlling deposition parameters it is possible to tune nanoparticle size [36] and chemistry up to the whole film mesostructure [37-39]. In order to explore the different hierarchical organizations attainable and their efficacy, nanostructured $TiO_2$ thin films were deposited by PLD, ablating a pure Ti target in a synthetic air background pressure in the 10 Pa to 100 Pa range. A decrease in film density of more than one order of magnitude with a consequent increase in surface area was observed (see below). In **Figure 1** we show SEM images of film cross sections at four different pressures. Film organization clearly changes passing from 10 Pa to 40 Pa. While at the lowest pressure a 500 nm thick, dense columnar film develops, increasing deposition pressure over 20 Pa causes cluster nucleation [34,35] and the assembly of these in a characteristic 'tree' shape. This behavior is attributed to the interplay of gas-phase nucleation, reduced impact kinetic energy and an effective shadowing effect with increasing pressure. These findings suggest the possibility to vary 'tree' height (i.e. film thickness), diameter and spacing by changing deposition parameters, hereby defining film density and surface area.

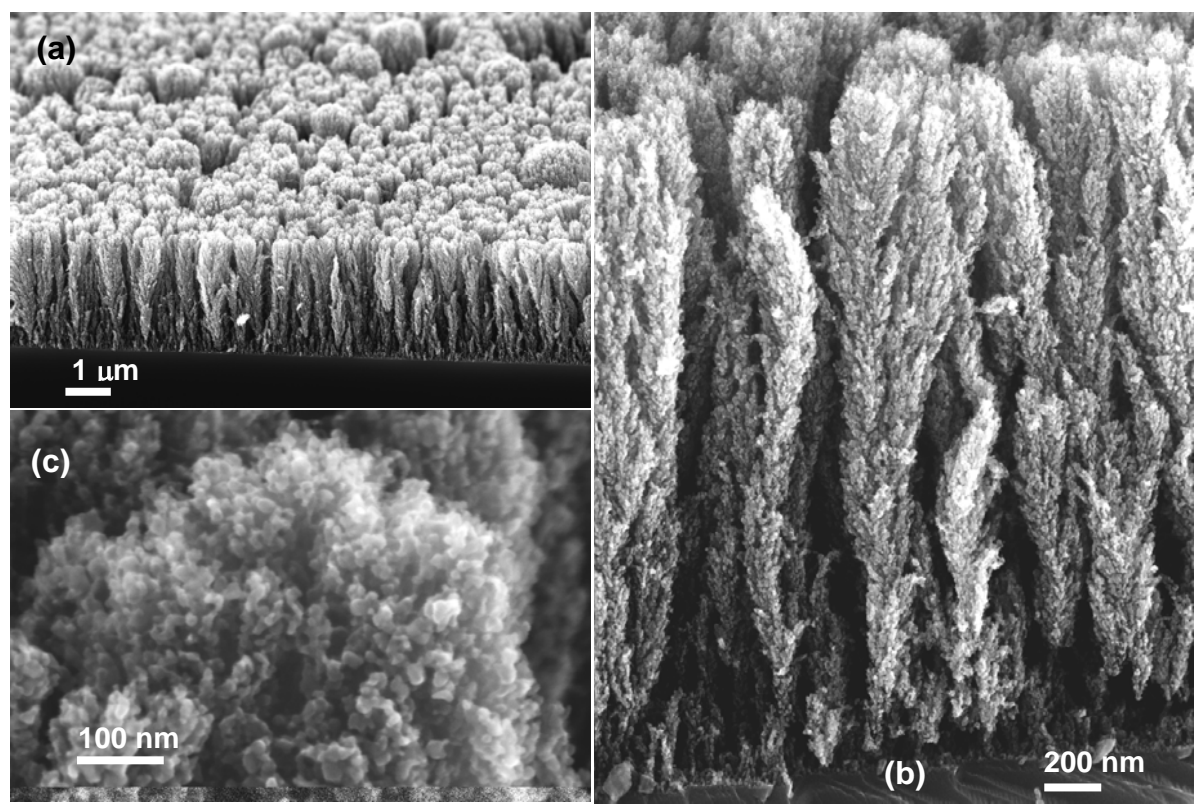

**Figure 2.** SEM images of the sample deposited at 40 Pa after one hour annealing in air at 400°C, showing the organization at different scales: a) the uniform "forest" like assembly; b) single "trees"; c) high resolution image a "tree" tip. This series of images clearly indicates that single $TiO_2$ nanoparticles are assembled in a three level hierachical structure.

The multiscale hierarchical organization of the film deposited at 40 Pa is clearly visible in SEM images at different magnification (**Figure 2**), showing the whole film assembly (the 'forest') and the single columnar structures (the 'trees') composed by nanometric particles (the 'leaves'). This same hierarchical organization is observed also in as deposited films revealing that annealing in air at 400°C for 1 hour does not significantly affect the film morphology at the meso- and nanoscale (while it affects the film crystalline structure, as discussed below). An increase in film porosity occurs when increasing the deposition pressure; since deposited mass was measured for each sample, we can calculate film density and porosity with respect to a film of the same total mass with bulk anatase density (100 μg over 1cm$^2$ would mean a 260 nm thick film). Sample deposited at 10 Pa shows about double the reference thickness indicating a 48% porosity. Increasing the deposition pressure yields a correspondent increase in porosity to 87% (20 Pa), 92% (40 Pa ) and over 94% (100 Pa), i.e. void fractions typical of aerogels [11-13].

In **Figure 3** the dependence of film density on deposition pressure is shown (density of bulk anatase is 3.895 g/cm$^3$). Moreover, in contrast to randomly organized aerogels, our films exhibit a well defined vertical meso-structure persistent up to very high void fractions. Increasing the pressure up to 100 Pa yields a transition to disordered and mechanically non-stable samples (see Fig.1d).

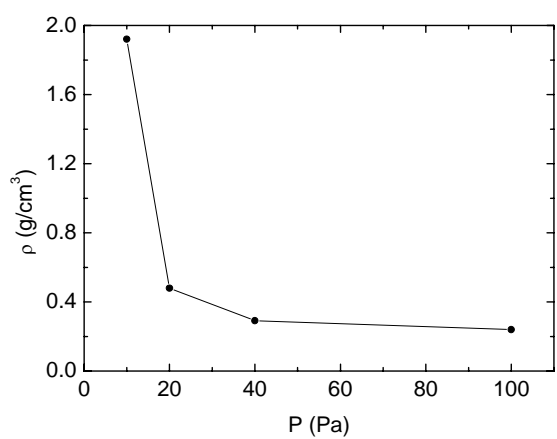

**Figure 3.** Nanostructured TiO$_2$ film density dependence on pressure. Increasing deposition pressure causes a sharp drop in film density due to gas-phase cluster nucleation, lower impact kinetic energy and enhanced shadowing effect. Bulk anatase density is 3.895 g/cm$^3$.

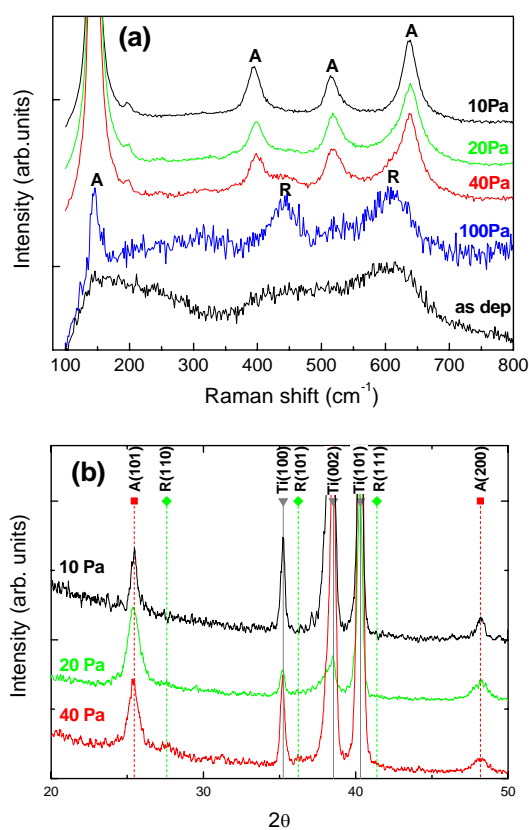

**Figure 4.** (a) Raman spectra of film deposited at increasing pressure and annealed in air at 400°C for one hour. A and R indicate anatase and rutile Raman peaks, respectively. A representative spectrum of the as-deposited films is shown. (b) XRD spectra of film deposited at increasing pressure and annealed in air at 400°C for one hour. A, R and Ti indicate anatase, rutile and titanium (substrate) reflexes, respectively.

Parallel to the change in morphology, a well defined evolution in the film structure occurs. Raman spectroscopy indicates that as deposited films are prevalently amorphous or disordered, while after annealing in air at 400 °C for 1 hour all spectra show crystalline signatures (**Figure 4a**) over an amorphous background. The dominant TiO$_2$ crystalline phase is anatase (Raman peaks at 143, 197,

399, 519 and 639 cm$^{-1}$), with an increasing rutile content (Raman peaks at 447 and 612 cm$^{-1}$) with increasing deposition pressure up to 100 Pa, where a predominantly rutile film is obtained. In this case, the intensity of the crystalline Raman peaks is very low and superimposed to large bands, suggesting a highly disordered structure and very fine grains (below 10 nm). From Raman spectra it is possible to identify the anatase-to-rutile relative content (AR=relative anatase percentage) in the film, as reported in ref. [40]. No quantitative absolute estimation can be obtained due to the presence of the amorphous phase. Obtained estimations confirm the tendency from a mainly anatase film for low pressure deposited films (10 Pa, AR=100%), to a film with a predominantly rutile crystalline content at high deposition pressure (100 Pa, AR=21.9%). In particular the sample deposited at 40 Pa shows an anatase relative concentration of 86.7%, very close to that of standard P25 powder. Moreover it is interesting to notice that 400 °C is well below the anatase-rutile transition temperature, suggesting the presence of rutile nanocrystals in as deposited samples, acting as nucleation centers even at moderate temperature. X-ray diffraction (XRD) spectra of annealed films confirm this evolution of the crystalline structure (**Figure 4b**). The XRD spectrum of the 100 Pa sample did not show any significant signature due to the reduced mass of the sample, nevertheless in ref. [32] we showed that in similar conditions a prevalently rutile phase is obtained, in agreement with the Raman spectrum of fig.4a. Parallel to the morphology and crystalline structure, also grain size exhibits an evolution with deposition pressure. By applying Scherrer's formula to both (101) and (200) anatase reflexes an average grain size was estimated of 20 nm for sample deposited at 10 Pa and of 10 nm for samples at 20 Pa and 40 Pa. In the latter a weak and broad signal from the (110) rutile reflex is present, allowing a rutile grain size estimate of about 6 nm. It has to be noted that a broader band is present under the anatase (101) peak, indicating also the presence of a distribution of much smaller, or very disordered, crystalline domains. This is also witnessed by blue shift and broadening of the anatase E$_g$ Raman peak at 143 cm$^{-1}$ [40], indicating a decreasing anatase crystalline domain size (ν = 143.6 cm$^{-1}$ and FWHM = 13 cm$^{-1}$ for 10 Pa vs. ν = 146.3 cm$^{-1}$ and FWHM = 16 cm$^{-1}$ for 100 Pa), in good agreement with the average grain size calculated from XRD spectra. As already mentioned, comparison of SEM images taken before and after annealing shows that no significant morphology modifications occur. Figure 2 shows that nanoparticle size roughly coincides with XRD crystalline domain size in the annealed samples. This evidences that annealing only results in structural ordering of the particles, while particle size growth and coalescence is hindered by the open morphology. These structural features reflect on film optical absorption. Band gap energies in the 3.2-3.5 eV range were found for as-deposited samples. These values are blue shifted with respect to crystalline bulk anatase (3.2 eV) and are consistent with highly disordered/amorphous titanium oxide films [7,41]. Upon annealing, band gaps red shift in the range 3.2-3.3 eV due to the increased crystalline order. The slight blue shift still present is attributed to the remnant amorphous fraction.

Photocatalytic efficacy was tested by mineralization (i.e. oxidation) of stearic acid. The reaction kinetics can be described by a modified Langmuir–Hinshelwood model in agreement with a surface reaction, as discussed, e.g. by Al-Ekabi and Serpone [42],

$$ln\frac{C_0}{C} + k(C_0 - C) = k_r K t \qquad (1)$$

where $C$ is the reactant concentration, $C_0$ the initial condition, $K$ is the absorption coefficient of the reactant, $k_r$ the reaction rate constant. Obviously, eq. (1) is the sum of zero-order and first-order rate equations, and their contribution to the overall reaction depends essentially on the initial concentration $C_0$. When $C_0$ is very small, eq. (1) reduces to the pseudo first order equation:

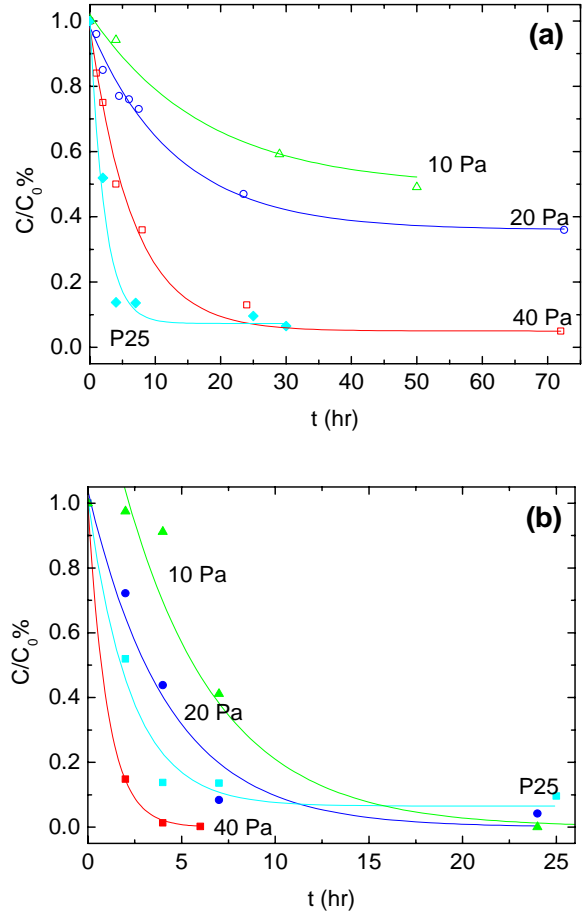

**Figure 5.** Stearic acid mineralization curves for samples as deposited (a) and annealed for 1 hour at 400°C (b). The mineralization curve for the P25 is included as reference. Solid lines are a guide to the eye.

$$ln\frac{C_0}{C} = k't \qquad (2)$$

where $k'$ is the apparent first-order rate constant. Abatement curves are plotted in **Figure 5** for the various deposition pressures, both for as deposited (a) and annealed (b) films. Samples deposited at 100 Pa were not studied because their scarce mechanical robustness did not allow for repeated tests.

First-order rate constants for the as deposited and annealed films are calculated and plotted in **Figure 6**. A sharp increase in degradation kinetics is evident going from 10 Pa to 40 Pa deposition pressure, corresponding to the above mentioned change in morphology. Noteworthy, as deposited 40 Pa sample shows an activity comparable with that of P25 powder. Upon annealing, photocatalytic activity sharply increases in all cases, in particular that of 20 Pa sample becomes comparable with P25's, while even faster kinetics than P25 is observed for the 40 Pa annealed sample. Moreover, complete decomposition of the organic compound is achieved after as little as 5 hours. In this case, plots of $ln(C/C_0)$ versus irradiation time (not shown) give a straight line for the 40 Pa film, while P25 powder and the 20 Pa film have an initial first order character followed by a zero order; the 10 Pa film is not simply definable since it shows a mixed rate behaviour. Annealing causes a net increase of activity of almost one order of magnitude for all samples. This effect may be ascribed to the increased crystalline order that has the beneficial effects of decreasing the number of defects, reducing the band gap and increasing the rutile content close to the that of P25. It has been proved, for nanostructured systems, that these occurrences, together with the reduced size of the single nanoparticles, enhance the photocatalytic performances by lowering the recombination rate of photogenerated electrons and holes (see refs. [19,20] and references therein).

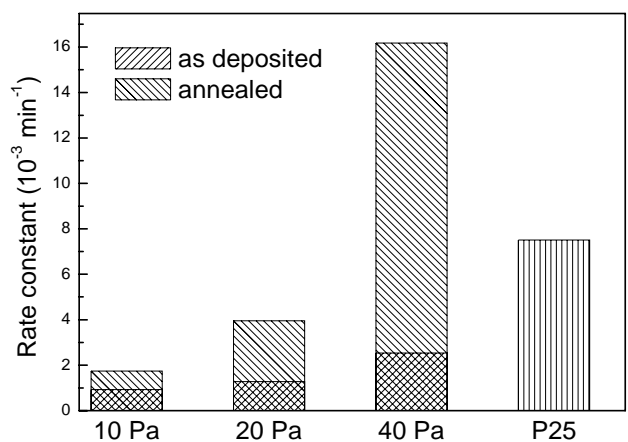

**Figure 6.** Comparison of the apparent first order time constant of hierarchically organized nanostructured TiO$_2$ prepared at different pressures before and after annealing and of P25.

We ascribe the observed kinetics of photocatalytic mineralization of stearic acid to the peculiar morphology of PLD grown samples. In fact, for given conditions, reaction rate is directly proportional to the fraction of catalytically active surface covered by stearic acid. Hence, the larger surface area of sample 40 Pa and its peculiar 3D morphology would allow for finely distribution of reactants on the surface and more readily accessible porosity than P25 powder. This interpretation is strengthened by direct comparison of the behavior of films deposited at 40 Pa and 20 Pa. After annealing these samples are similar in A/R ratio and grain size (with band gap slightly smaller, i.e. favorable, for the 20 Pa sample), nevertheless the film deposited at 40 Pa exhibits a reaction rate more than two times faster.

**Figure 7** shows that the obtained structure favors uniform wetting of the surface by the stearic acid and penetration in the spaces between adjacent columns, since no clumps of stearic acid are left on the finely segmented surface of the film (as instead observed in compact films deposited at 10 Pa, not shown). The smaller TiO$_2$ crystal size means larger porosity and larger surface area per gram of the sample, even though absolute values of the specific surface depend on the particular particle assembly. BET measurements indicate a specific surface area for the 40 Pa film of about 300 m$^2$/g, while specific surface area of standard Degussa P25 is of the order of 50 m$^2$/g. This contributes to the enhanced photocatalytic activity of the nanostructured TiO$_2$ film. On the other hand, in sample deposited at 10 Pa the active area is essentially the geometric 2D surface area, thus the same quantity of stearic acid accumulates on the surface, yielding to a shortage of oxygen in the first stage of the reaction. After some time, holes are generated in the capping layer of stearic acid allowing ambient oxygen to diffuse faster to the surface, increasing the reaction rate. This is consistent with what observed by Yu [19,20] for hierarchical macro-mesoporous titania obtained via hydrothermal preparation.

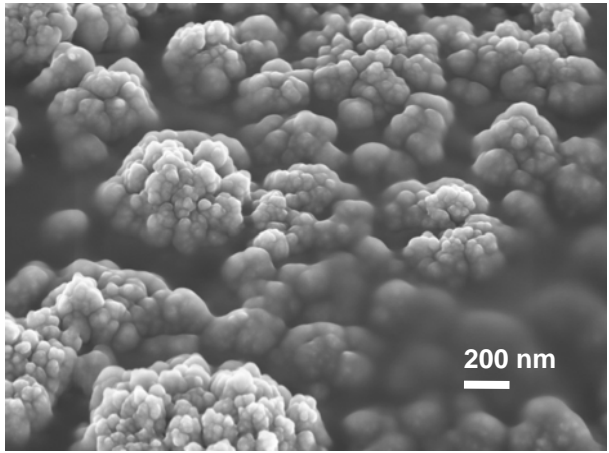

**Figure 7.** SEM image of sample deposited at 40 Pa and annealed in air for 1 hour at 400°C after deposition of stearic acid, showing uniform wetting of the surface by the stearic acid.

## 4. Conclusions

A number of studies have been devoted to investigating the correlation among structural/ morphological properties and the photocatalytic activity of titanium oxide films. As stated in the introduction, a hierarchically ordered assembly of nanoparticles is ideal for a number of reasons in photocatalytic and photovoltaic applications. In this study, we demonstrated the possibility to obtain, by proper control of pulsed laser deposition parameters, $TiO_2$ films with different organizations, density, particle size and specific surface area. Enhanced photocatalytic activity (stearic acid mineralization) was demonstrated with respect to standard P25 powder even for as deposited samples. Our 'forest of trees' arrangement favours deep penetration of organic molecules through the film and very large surface area for light absorption and oxygen interaction. This is crucial, as the functional part of the light-absorbing material is the surface in direct contact with the molecules. Permeation of the organic compound through the film thickness is reputed also to be fundamental for complete mineralization. This is interesting for $TiO_2$-based applications, since PLD allows deposition on inexpensive substrates, even plastics, kept at room temperature. Furthermore, single-step deposition of variable density films with good adhesion to the substrate is possible by PLD in a variable background pressure, an interesting perspective both for photocatalytic and advanced photovoltaic application, substituting the time consuming deposition of different layers and long lasting annealing steps.

**Acknowledgements**
The authors wish to thank L. Lietti (Department of Energy, Politecnico di Milano) for BET measurements and A. Lucotti (Department of Chemistry, Materials and Chemical Engineering, Politecnico di Milano) for diffuse reflectance spectroscopy measurements.